\documentclass[
    ,final            
    ,sort&compress
  ]
  {aipproc}

\layoutstyle{8x11double}

\usepackage{natbib}

\newcommand{\Msun}{\mbox{$\mathrm{M}_{\odot}$~}}
\newcommand{\Rsun}{\mbox{$\mathrm{R}_{\odot}$~}}
\newcommand{\Msunpyr}{\mbox{$\mathrm{M}_{\odot}/{\rm yr}$~}}

\begin{document}

\title{Can low metallicity binaries avoid merging?}


\author{S.~E. de Mink}{
  address={Astronomical Institute Utrecht,
              Princetonplein 5, NL-3584 CC Utrecht, The Netherlands,
              S.E.deMink@astro.uu.nl}
}
\author{M. Cottaar}{}
\author{O.~R. Pols}{}

\begin{abstract}
  Rapid mass transfer in a binary system can drive the accreting star
  out of thermal equilibrium, causing it to expand. This can lead to a
  contact system, strong mass loss from the system and possibly
  merging of the two stars. In low metallicity stars the timescale for
  heat transport is shorter due to the lower opacity. The accreting
  star can therefore restore thermal equilibrium more quickly and
  possibly avoid contact.

  We investigate the effect of accretion onto main sequence stars with
  radiative envelopes with different metallicities.  We find that a
  low metallicity ($Z<10^{-3}$) $4\Msun$ star can endure a 10 to 30
  times higher accretion rate before it reaches a certain radius than
  a star at solar metallicity.  This could imply that up to two times
  fewer systems come into contact during rapid mass transfer when we
  compare low metallicity. This factor is uncertain due to the unknown
  distribution of binary parameters and the dependence of the mass
  transfer timescale on metallicity.
  In a forthcoming paper we will present analytic fits to models of
  accreting stars at various metallicities intended for the use in
  population synthesis models.

\end{abstract}

\maketitle


\section{Introduction}
  
The majority of stars are found in binaries, many of which can
interact, for example by exchanging mass, resulting in an evolution
very distinct from isolated stars. Although the fraction of stars in
binaries might be different for earlier generations of stars, formed
in metal-poor environments, they are certainly worth a systematic
study.  In this work we discuss the effect of accretion onto main
sequence stars as function of metallicity. In a second contribution to
these proceedings we discuss how the ranges for different cases of
mass transfer depend on metallicity \citep{caseABC07}.

Mass transfer takes place when one of the stars exceeds a certain
critical radius, the Roche lobe radius. The first phase of mass
transfer is usually so fast that the accreting star is driven out of
thermal equilibrium and expands \citep{Benson70,Yungelson73},
potentially so much that the two stars come into contact.  
Contact binaries are not well understood, but they probably involve 
strong mass loss from the system and merging of the two stars.
  
Instead of using full binary evolution models, we choose to reduce the
large parameter space of binaries by studying the behavior of models
of isolated stars under controlled conditions.  We follow the approach
of \citet{Kippenhahn+Meyer77} and \citet{Neo+ea77}, who studied the
evolution of the radius of accreting main sequence stars. We extend
their work, to study the effect of metallicity, using up-to-date input
physics.

Here we present the results of an exploratory study and its possible
implications for binaries at low metallicity. In a forth coming paper
we will present analytic fits to a finer grid of models, intended for
the use in population synthesis models. Expansion due to thermal
timescale mass transfer is commonly neglected or taken into account
using a simple approximate criterion. Our fits will provide an
improvement, which is easy to implement.

   \begin{figure}
     \includegraphics[ angle=-90,
       width=\columnwidth]{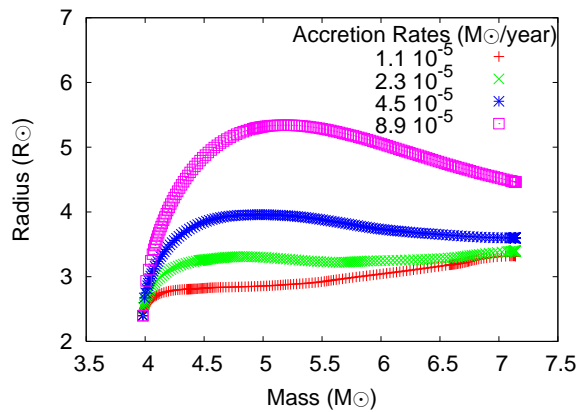} \caption{Radius versus mass
       of an initially 4 \Msun star at solar metallicity accreting with
       different accretion rates. \label{R_vs_m} }
   \end{figure}

\section{Evolution code}  

We use the fully implicit stellar evolution code \texttt{STARS},
originally developed by Peter Eggleton \citep{Eggleton71, Eggleton72,
  Pols+ea95}, updated with the latest opacity tables
\citep{Eldridge+Tout04, OPAL96,Ferguson+Alexander05}.  For the
hydrogen and helium abundance we assume a linear relation with $Z$: $X
= 0.76 -3.0 Z$ and $Y = 0.24 + 2.0 Z$.  The abundances of the heavier
elements are assumed to scale to solar and meteoric abundances
\citep{Anders+Grevesse89} so that they are consistent with the opacity
tables.  A mixing length ratio $l/H_\mathrm{p}=2.0$ is
assumed. Convective overshooting is taken into account as in
\citet{Schroeder+ea97} with a free parameter $\delta_\mathrm{ov}=0.12$
calibrated against accurate stellar data from non-interacting binaries
\citep{Pols+ea97}.

We assume that material is accreted in a spherically symmetric way at
a constant rate onto a zero age main sequence star with a composition
and specific entropy equal to the surface value.

\section{Results}

   Figure~\ref{R_vs_m} shows as an example the radius versus mass $M$,
   which increases linearly with time, of an initially $4\Msun$ star
   at solar metallicity ($Z=0.02$) accreting at different rates.  At
   an accretion rate of $8.9~10^{-5} \Msunpyr$ the star expands to
   more than twice its original radius after accreting $1\Msun$ before
   it shrinks again towards its thermal equilibrium radius $R_{\rm
   eq}(M)$. If the accretion rate is low, for example $1.1~10^{-5}
   \Msunpyr$, the radius of the star remains close to its equilibrium
   radius. The equilibrium radius $R_{\rm eq}(M)$ increases with the
   mass of the star. We define $R_{\rm max}$, as the radius of the
   star when $R/R_{\rm eq}(M)$ is at maximum, so that it represents
   the radius at the moment the star is driven furthest out of thermal
   equilibrium.

\begin{figure}
     \includegraphics[angle=-90, width=\columnwidth]{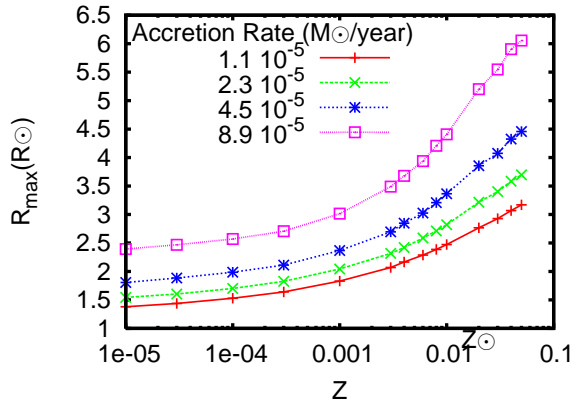}
       \caption{The maximum radius during accretion (for definition see
        text) is plotted versus metallicity for a 4\Msun star accreting at
	 different rates.\label{rmax_vs_Z} }
    \end{figure}

   The effect of the metallicity on $R_{\rm max}$ is shown in
   Figure~\ref{rmax_vs_Z}. The maximum radius increases by more than a
   factor of two if we compare low metallicity ($Z<10^{-3}$) to solar
   metallicity. Especially in the range $Z>10^{-3}$ the effect becomes
   important.

   In order to expand to a radius of 2.5\Rsun a low metallicity star
   needs to accrete approximately 8.9\Msunpyr, while a solar
   metallicity will expand to this radius when it accretes
   1.1\Msunpyr, see Fig.~\ref{rmax_vs_Z}.  

   From our models we estimate that a star can endure a 10 to 30 times
   higher accretion rate, comparing low to solar metallicity before
   its radius exceeds a certain critical radius, say the Roche lobe
   radius.

   This effect can be understood by considering the timescale for heat
   transport $\tau_{\rm h}$ \citep[][eq. 10]{Neo+ea77}
    \[
    \tau_{\rm h} = \frac{3 C_{\rm p} \kappa \rho^2 H_{\rm p}^2 }{4 a c
      T^3}
   \] 
   over one scale height of pressure $ H_{\rm p}$, where $C_{\rm p}$
   and $\kappa$ denote the specific heat and opacity.
   In metal rich stars, heavy elements like C, O, N and Fe provide an
   important contribution to the opacity, increasing the timescale for
   heat transport.  At low metallicity stars can radiate away any
   excess energy on a shorter timescale, to restore thermal
   equilibrium.

\section{Implications}

   In this paragraph we will speculate on the possible implications of
   our findings for binary stars.
   In binaries with nearly equal masses contact during rapid mass
   transfer is avoided, as the time scale for mass transfer, which is
   of the order of the thermal time scale of the donor, is of the same
   order as the time scale on which the accreting star can restore its
   thermal equilibrium. If the mass ratio $q = M_a/M_d $ is smaller
   than a certain critical value $q_{\rm crit}$ the systems do get
   into contact.  For solar metallicity $q_{\rm crit} \approx 0.56$
   \cite{Nelson+Eggleton01}, for close systems in the mass range of
   our models.  From our results we can roughly estimate how $q_{\rm
   crit}$ changes with metallicity.

   If we assume that the mass transfer rate in a binary can be
   approximated by the mass of the donor $M_{\rm d}$ over the thermal
   timescale of the donor star $\tau_{KH} = G M^2/RL \approx
   M^{\alpha}$, where $\alpha =2.7..2.9$, then a 10 to 30 times higher
   critical accretion rate corresponds to a donor which is at least
   twice as massive. Therefore $q_{\rm crit}$ will be a factor 2
   smaller at low metallicity.
   
   If we make the commonly used assumption that the initial
   distribution of mass ratios is flat in q, and that the
   distributions do not depend on metallicity and if we assume that
   the mass transfer rate does not depend on the metallicity, nor the
   binary parameters, then our result implies that the number of
   systems that come into contact during rapid mass transfer at low
   metallicity is half of that at solar metallicity.

\section{Conclusions}

   We find that a $4\Msun$ low metallicity star ($Z<10^{-3}$) can endure
   a 10 to 30 times higher accretion rate before it expands to a certain
   radius compared to solar metallicity stars.

   This suggests that at low metallicity fewer binaries come into
   contact during rapid mass transfer. A rough estimate based on very
   simplified assumptions indicate that the effect could be up to a
   factor 2, i.e. only half as many may binaries evolve into contact
   at low metallicity(Z < 10{-3}) compared to solar metallicity.

   This factor is uncertain and probably only an upper limit as it
   depends on the initial distribution of binary parameters, on how
   the typical mass transfer rate depends on metallicity \citep[it is
   probably higher at low metallicity, e.g.][]{Langer_ea00} and the
   binary parameters, on the specific entropy of the accreted material
   and on the efficiency of mass transfer \citep[see for
   example][]{DeMink+ea07}.

   In a forthcoming paper we will present analytic fits to our models
   intended for the use in population synthesis models.


\begin{theacknowledgments}
We would like to thank Peter Eggleton for providing his stellar
evolution code, John Eldridge for the opacity tables and Rob Izzard
for interesting discussions and suggestions.
\end{theacknowledgments}

\bibliographystyle{aipproc}   

\bibliography{SF07_acc}

\IfFileExists{\jobname.bbl}{}
 {\typeout{}
  \typeout{******************************************}
  \typeout{** Please run "bibtex \jobname" to optain}
  \typeout{** the bibliography and then re-run LaTeX}
  \typeout{** twice to fix the references!}
  \typeout{******************************************}
  \typeout{}
 }

\end{document}